\documentclass[useAMS,usenatbib]{mn2e}

\def\msun{\hbox{M$_{\odot}$}~}

\def\rsun{R$_{\odot}$}

\usepackage{graphicx}  
 \usepackage{times}
  \usepackage{lscape}
  \usepackage{rotating}
  \usepackage{txfonts}

\title[The eclipsing Algol OGLE-LMC-DPV-065]{ On the long cycle variability of the Algol OGLE-LMC-DPV-065 and its stellar, orbital and disk parameters}
\author[Mennickent et al. ]
  {R.E. Mennickent$^{1}$\thanks{E-mail: rmennick@udec.cl. Based on ESO proposal 092.D-0385(A) and CNTAC 2014B-13.},  
  M. Cabezas$^{1,2}$,
G.  Djura{\v s}evi{\'c}$^{3,4}$,
  T. Rivinius$^{5}$,
  P. Hadrava$^{2}$,
  R. Poleski$^{6,7}$, \newauthor 
     I. Soszy{\'n}ski$^{7}$,
     L. Celed\'on$^{1}$, 
     N. Astudillo-Defru$^{1,8}$,
     A. Raj$^{9}$, 
     J. G. Fern\'andez-Trincado$^{1,10,11}$, \newauthor
     L. Schmidtobreick$^{5}$, C. Tappert$^{12}$,
     V. Neustroev$^{13}$,
     I. Porritt$^{14}$  \\
  $^1$Universidad de Concepci\'on, Departamento de Astronom\'{\i}a,
      Casilla 160-C, Concepci\'on, Chile\\
      $^{2}$Astronomical Institute of the Academy of Sciences of the Czech Republic,  Bo\v{c}n\'i II 1401/1, Prague, 141 00, Czech Republic\\
       $^{3}$Astronomical Observatory, Volgina 7, 11060 Belgrade 38, Serbia\\
   $^{4}$ Isaac Newton Institute of Chile, Yugoslavia Branch\\
      $^{5}$ European Southern Observatory\\
      $^{6}$ Department of Astronomy, Ohio State University, 140 W. 18th Ave., Columbus, OH 43210, USA\\
       $^{7}$ Warsaw University Observatory, Al. Ujazdowskie 4, 00-478 Warszawa, Poland \\ 
     $^{8}$   Departamento de Matem\'atica y F\'isica Aplicadas, Universidad
Cat\'olica de la Sant\'isima Concepci\'on, Alonso de Rivera 2850,
Concepci\'on, Chile\\
       $^{9}$  Indian Institute of Astrophysics, II Block Koramangala, Bangalore 560034, India \\
         $^{10}$  Institut Utinam, CNRS UMR6213, Univ. Bourgogne Franche-Comt\'e, OSU THETA, 
       Observatorie de Besan\c{c}on, BP 1615, 25010 Besan\c{c}on Cedex, France \\ 
        $^{11}$ Instituto de Astronom\'{\i}a y Ciencias Planetarias, Universidad de Atacama, Copayapu 485, Copiap\'o, Chile \\ 
     $^{12}$ Instituto de F{\'{i}}sica y Astronom{\'{i}}a, Universidad de Valpara{\'{i}}so, Chile \\
     $^{13}$ Astronomy research unit, P.O. Box 3000 90014 University of Oulu, Finland \\
       $^{14}$ Turitea Observatory, Palmerston North, New Zealand\\
         }
\date{}

\begin{document}


\maketitle 

\begin{abstract} 

OGLE-LMC-DPV-065 is an interacting binary whose  double-hump long photometric cycle remains hitherto unexplained. We analyze photometric time series available in archive datasets spanning 124 years and present the analysis of new high-resolution spectra.
A refined orbital period is found of 10\fd0316267 $\pm$ 0\fd0000056 without any evidence of variability.
In spite of this constancy, small but significant changes in timings of the secondary eclipse are detected. 
We show that the long period continuously decreases from 350 to 218 days during  13 years, then remains almost constant for about  10 years. 
Our study of radial velocities indicates a circular orbit for the binary and yields a mass ratio of 0.203  $\pm$ 0.001. From the analysis of the orbital light curve 
we find that the system contains 13.8 and 2.81 \msun\ stars of radii 8.8 and 12.6 \rsun\ and
absolute bolometric magnitudes -6.4 and -3.0, respectively. The orbit semi-major axis is 49.9 \rsun\ and the stellar temperatures are 25460 K and 9825 K. We find evidence for 
an optically and geometrically thick disk around the hotter star. According to our model, the disk has a radius of  25 \rsun, central and outer vertical thickness of 1.6 \rsun\ and 3.5 \rsun, and temperature of 9380 K at its outer edge. Two shock regions located at roughly opposite parts of the outer disk rim can explain the light curves asymmetries. The system is a member of the double periodic variables 
and its relatively high-mass and long photometric cycle make it similar in some aspects to $\beta$ Lyrae.

\end{abstract}


\section{Introduction}

Stellar magnetic dynamos are relatively common in solar type stars, and magnetic activity in binaries containing GK dwarfs  is well documented in the RS\,CVn systems \citep{1989SSRv...50..219H}. The situation in Algol-type variables is less clear. Algols are semi-detached binaries with intermediate mass components, where the less massive star, dubbed secondary or donor, is more evolved than the more massive star, named gainer or primary. Some authors have proposed that orbital period changes  observed in some close binaries might be driven by magnetic cycles through the Applegate (1992) mechanism; the angular momentum of the star and the binary is redistributed during the magnetic cycle producing the observed orbital period changes \citep{1998MNRAS.296..893L, 1999A&A...349..887L}. Further studies indicate that the presence of a dynamo may modulate the mass transfer rate in Algol systems, leading to a characteristic impact of the dynamo cycle on the system luminosity \citep{1989SSRv...50..311B,  2004MNRAS.352..416M}. 
In this context the existence of a group of hot Algols showing a long photometric cycle lasting on average about 33 times the orbital period might be relevant, since this variability  has been recently interpreted in terms of a magnetic dynamo \citep{2017A&A...602A.109S}. If this hypothesis turns to be correct, one may deduce that the stellar dynamo is also active in the hot, rapidly rotating (orbitally synchronized) A-type giants in some semidetached Algols.  In fact, for the Algol binary V393\,Sco indirect evidence for magnetism in the secondary star has been deduced from the presence of chromospheric emission lines \citep{2018PASP..130i4203M}. These authors note that the spin-up of the donor during mass-transfer stage increases its dynamo number, likely enhancing the probability of occurrence of a magnetic dynamo at the semi-detached stage. 

The aforementioned group of hot Algols showing long photometric cycles additional to their orbital variability is named Double Periodic Variables \citep[DPVs,][]{2003A&A...399L..47M, 2016MNRAS.455.1728M, 2010AcA....60..179P, 2013AcA....63..323P, 2017SerAJ.194....1M}.
DPVs are semidetached binaries  typically consisting of a A/F/G giant star  filling its Roche lobe and transferring mass onto a B-type primary surrounded by a circumprimary disk. Among Galactic DPVs, one famous example is $\beta$ Lyrae \citep{1989SSRv...50...35G, 1996A&A...312..879H, 2002AN....323...87H}.

Few extragalactic DPVs have been studied at some detail. Among them, the case of
OGLE-LMC-DPV-065 (OGLE05200407-6936391; R.A.$_{2000}$ = 05:20:04.07, Dec.$_{2000}$ = -69:36:39.1)
is notable, since it is one of the brightest DPVs in the LMC
($V$ = 14.74, $B-V$ = -0.07), and shows a remarkable change in the long period from 350 to 210 days in 15 years
that clearly stands out among the rest of the DPVs. In addition, the system is eclipsing, with a 1.4 mag deep
main eclipse and a comparatively large amplitude of the long  cycle of 0.3 mag, in the $I$ band.
To date, there is no other DPV with such a remarkable change in the long cycle.
The orbital period has been reported as ${\rm P_o}$ = 10\fd031645  $\pm$ 0.000033 \citep{2010AcA....60..179P}.

The above credentials make  OGLE-LMC-DPV-065 an ideal target for a deeper study. If the long cycle is related to changes in the mass transfer driven by a magnetic dynamo,
it might show up in spectroscopic and photometric signatures during the long cycle. For this reason we conducted a long-term spectroscopic monitoring of this target with UVES at the VLT (Sec. 2.3). For the sake of order and clarity we have divided our work in two parts. In this first paper we analyze the available photometric time series making use of archive data, present our new high-resolution spectroscopic observations, calculate the system and orbital parameters  and provide a solution for the stellar radius, mass, luminosity, surface gravity along with a
characterization of the accretion disk. In a second forthcoming paper we will provide an analysis of the spectroscopic changes during the long cycle and a study of the evolutionary stage of the binary. {We notice that a short and preliminary spectroscopic study of this object based on the data presented in this paper has been presented in a recent conference \citep{Cabezas}.}

\section{Observations and methods}

\subsection{Photometric observations}


We included OGLE-II  \citep{2005AcA....55...43S}\footnote{http://ogledb.astrouw.edu.pl/$\sim$ogle/photdb/} and OGLE-III/IV data\footnote{ 
OGLE-III/IV data kindly provided by the OGLE team.}. The OGLE-IV project is described by \citet{2015AcA....65....1U}.
Poleski et al. (2010) published the OGLE-II and OGLE-III  $I$-band\footnote{ftp://ftp.astrouw.edu.pl/ogle/ogle3/OIII-CVS/lmc/dpv/phot/I/OGLE-LMC-DPV-065.dat}
and $V$-band\footnote{ftp://ftp.astrouw.edu.pl/ogle/ogle3/OIII-CVS/lmc/dpv/phot/V/OGLE-LMC-DPV-065.dat}
data of  this star.
We also considered 460 $B$ magnitudes from the Digitalized Harvard plates (DASCH project)\footnote{http://dasch.rc.fas.harvard.edu/project.php} covering 59.4 years, since August 1893 to January 1953.
In addition, we obtained new photometry  with the CTIO 1.3m telescope operated by the SMARTS consortium in service mode between November 2014 and October 2015,
with the ANDICAM camera and filters $V$ and $I$.
Another data set was collected by Ian Porritt in Turitea Observatory, New Zealand, with the 0.36-meter Meade telescope and a yellow filter. These new data were reduced in the usual way, 
removing bias and performing flat field corrections in the images and calculating differential magnitudes with respect to constant comparison stars. 
Finally, 664 $V$-band magnitudes were included from the ASAS-SN catalogue.
The photometric observations analyzed in this paper amount to 3099 data points, cover 124 years and are summarized in Table\,1. 

\begin{table}
\centering
 \caption{Summary of photometric observations considered in this paper. 
 The number of measurements, starting and ending times for the series, average magnitude and standard deviation (in magnitudes) are given.
 Single point uncertainties in the $I$-band and $V$-band for OGLE data are between 4 and 6 mmag. The zero point for HJD is 2\,400\,000. See comment on the average magnitudes in the text.}
 \begin{tabular}{@{}lrrcccc@{}}
 \hline
Source &N &HJD$_{start}$ &HJD$_{end}$ &mag &std. &band \\
\hline
DASCH&460&12697.8482 & 34399.4995 &14.996 &0.219 &$B$\\
OGLE-II&915&50455.6744 & 51873.7744&14.898 &0.218 &$I$ \\
OGLE-III&504&52123.9345 & 54953.5268&14.907 &0.246 &$I$ \\
OGLE-IV & 73 &55326.4931 & 57710.7482 &14.901 &0.302& $I$ \\
CTIO &97&56964.7927 & 57327.7354&14.901 &0.317 &$I$ \\
OGLE-II&95 &50467.7237 & 51631.5633 &14.908 &0.244 &$V$\\
OGLE-III&90 &52990.6851 & 54948.4703 &14.929 &0.245 &$V$\\
Turitea &106&56342.9193 & 56467.8381 &14.918 &0.368 &$y$ \\
ASAS-SN &664 &56789.4535 & 57974.8870 &14.918 &0.116 &$V$\\ 
CTIO &95&56964.7954 & 57327.7381 &14.918 &0.330 &$V$\\ 
 \hline
\end{tabular}
\end{table}


\subsection{Light curve disentangling}

We separated the light curve into long- and a short-period components. For that
we used an algorithm especially designed to disentangle multi-periodic
light curves through the analysis of their Fourier component
amplitudes. The method is described in Mennickent et al. (2012) and a short summary is given here. The main frequency
$f_1$  is found with a period searching algorithm, this is usually the orbital frequency.
A least-square fit is then applied to
the light curve with a fitting function consisting of a sum
of sine functions representing
the main frequency and their additional significant harmonics. 
Afterwards the residuals are inspected for a new periodicity
$f_2$. This new periodicity (the long cycle in the case of DPVs) and their
harmonics are included in the new fitting procedure. Finally, we
obtain the light curve represented by  a sum
of Fourier components of frequency $f_1$ and $f_2$ and their harmonics.
Data residuals with respect to the second and first theoretical
light curves are the photometric series representing the orbital and long cycle,
respectively.

\subsection{Spectroscopic observations}

We were granted 25 hours for spectroscopic observations of the target 
with the ESO Ultraviolet and Visual Echelle Spectrograph UVES at the Kueyen telescope in the Paranal Observatory 
in service mode. This is a cross-dispersed echelle spectrograph designed to operate with high efficiency from the atmospheric cut-off at 300 nm 
to the long- wavelength limit of the CCD detectors (about 1100 nm). The light beam is split into two arms, UV-Blue and Visual-Red, within the instrument. The two arms can be operated separately or in parallel with a dichroic beam splitter. The instrument provides
accurate calibration of the wavelength scale down to an accuracy of at least 50 m/s.

With the aim of covering both the orbital as well as the long cycle 
27 spectra were secured between October 1, 2013 and February 1, 2015 with the dichroic\#2 mode
in the ranges 3760$-$4985, 5700$-$7520 and 7665$-$9445 \AA. The slit widths of 0\farcs9 at the blue channel and 0\farcs8 at the red channels 
provided a resolving power of 50\,000 and 55\,000, respectively. 
The object was observed at typical airmass 1.4 and with 3000\,s exposure time per single exposure.
A typical S/N of 65 was achieved at 480 nm. 
A summary of the spectroscopic observations is given in Table\,7.






\begin{figure*}
\scalebox{1}[1]{\includegraphics[angle=0,width=17cm]{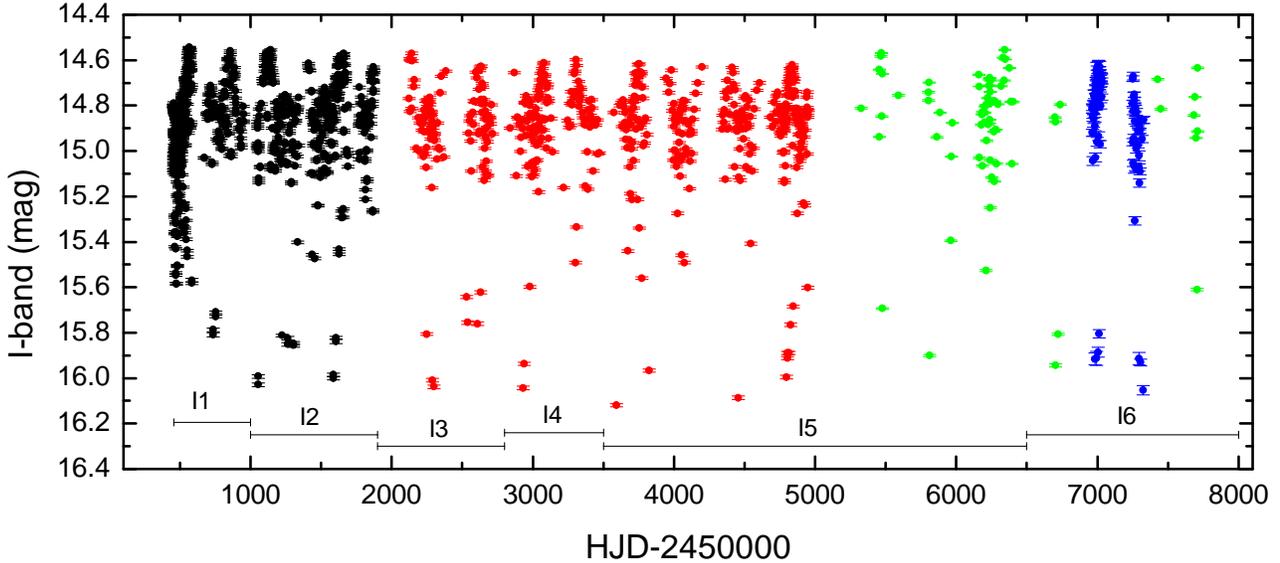}}
\caption{Light curve for OGLE\,II (black), III (red), and IV (green) datasets along with CTIO (blue) data. The upper envelope is produced by the long-period variability. The data 
points fainter than about 15.2 mag are taken during primary eclipse.  The data intervals given in Table\,4 are also shown. }
  \label{ILC}
\end{figure*}

\begin{figure}
\scalebox{1}[1]{\includegraphics[angle=0,width=8cm]{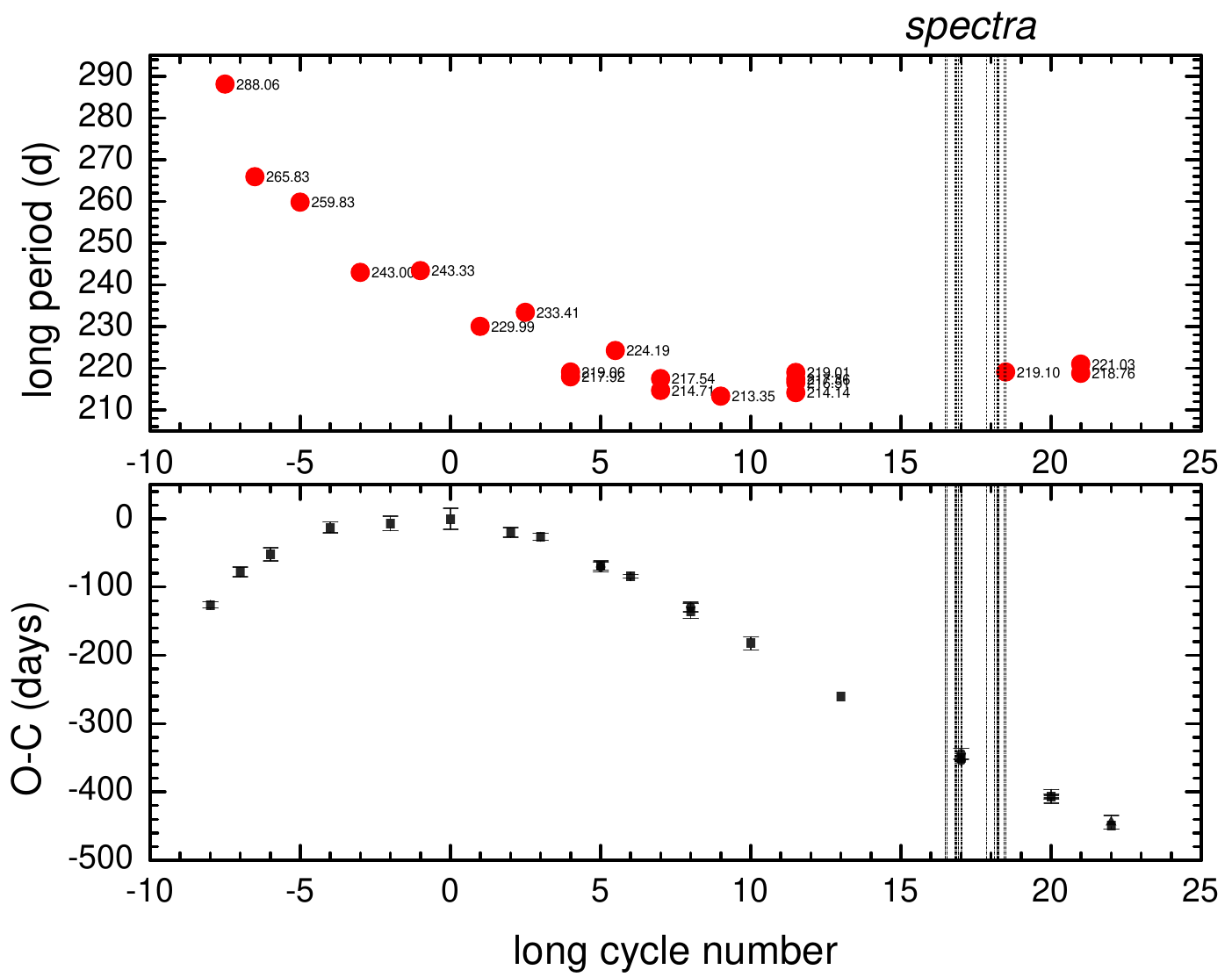}}
\caption{(up): Long period versus long cycle number. (down): Observed minus calculated times of the long cycle maximum versus 
long cycle number. Data are based on a test period of 240 days and are given in Table\,3. Vertical dashed lines indicate spectroscopic observation times.}
  \label{OCgraph}
\end{figure}

%
%
%
%

\section{Data analysis}
 
\subsection{The orbital period}

Zero points can be an issue when combining photometry obtained at different sites with different detectors, filters and sky conditions. 
We have shifted OGLE\,IV, ASAS-SN, Turitea, DASCH and CTIO magnitudes to the OGLE-II and OGLE-III  averages before performing the analysis described in this paper.
In Table\,1 we provide only the original magnitude average for OGLE-II and OGLE-III data and also for the DASCH $B$-band magnitude.
The light curve in the $I$-band is shown in Fig.\,1.

We conducted a search for the orbital period using standard methodologies: eclipse times were measured interactively in the light curve with the computer cursor and
a straight line fit was performed with the measured (epoch, time) pairs; the resulting slope gave the orbital period and their error. 
We also used the \texttt{Period04} program, that calculates errors based on a Monte Carlo technique (Lenz \& Breger 2005). The $I$-band residuals
were obtained after removing the long-term cycle (see next section).
The periodicities found in different datasets are given in Table\,2.
We can see that the data are consistent with a constant orbital period; we find the following ephemerides for the main eclipse:\\

\begin{equation}
HJD_{min} =  2\,450\,453.025 + (10\fd0316267 \pm 0\fd0000056)\,E. 
\end{equation}

\begin{table}
\centering
 \caption{Summary of the search for the orbital period $P_o$. }
 \begin{tabular}{@{}ccr@{}}
 \hline
$P_o$ & Method & note \\
\hline
10\fd0316450 $\pm$ 0\fd0000330 &- &Poleski et al. (2010)\\
10\fd0317800 $\pm$ 0\fd0000780 &Eclipse timings & $I$ band \\
10\fd0316259 $\pm$ 0.0000230 &Period04 &$I$ band \\
10\fd0316173 $\pm$ 0.0000128 &Period04 &$I$ residuals \\
10\fd0316257 $\pm$ 0.0000066 &Period04 &$B$ and $I$ bands\\ \hline
 10\fd0316267 $\pm$ 0\fd0000056 & &Weighted mean \\
\hline
\end{tabular}
\end{table}

\subsection{The orbital and long cycle light curves}

Maxima of the long cycle were measured directly from the light curves and compared with the ephemerides for
a 240 day test period, as reported in Table\,3. The observed minus calculated ($O-C$) diagram, constructed with the observed times of maxima ($HJD_O$) and the predicted 
ones ($HJD_C$), shows that the  long cycle decreased at the beginning of the observations
then remained more or less constant during about  14 cycles (Fig.\,2). We notice that considering the MACHO data analyzed by \citet[][HJD: 2448900-2451500, not included in this paper]{2005MNRAS.357.1219M}, which is previous to the OGLE data reported here, the long period has decreased from about 350 to 218 days continuously during about  13 years, before entering in a phase of almost constant period, that lasted for slightly more than  10 yr.

At every epoch we defined a local long cycle period $P_l$, subtracting the observed maximum timing from the previous one and 
dividing by the number of elapsed times, as given in Table\,3. 
 After inspection of Fig.\,2 we choose six data  intervals characterized by a more or less constant long cycle  and large number of observations (Table\,4). 
 This procedure allowed to apply the disentangling to every data block considering the variability of the long cycle. 
 The resulting disentangled light curves are shown in Fig.\,3, they reveal that the long cycle is double-humped and that it shape remains relatively constant. 
 In addition, the orbital light curve shows a small but significant variability (Figs.\,4 and 5):
(i) on the 5th interval between $HJD$ 2\,455\,804 and 2\,456\,405
 the system is brighter at quadratures, and produces  larger scatter in the long cycle light curve, (ii) on the
 first interval the main eclipse seems to be shallower, (iii) significant variability is observed during secondary eclipse; the secondary eclipse seems to occur earlier in interval 1 than in interval 2, and (iv) the shape of the eclipses vary minimally during the maximum, the minimum and the secondary maximum of the long cycle, perhaps the egress of the main eclipse around phases 0.1-0.2 is shallower during the low stage.  The changes in timing of minima during the secondary eclipse might indicate changes in the photo-center   
 of the eclipsed or eclipsing source, or changes in circumstellar matter or the donor hemisphere facing the gainer. An unseen/undetected body that dynamically affects the photo-center is another possibility.
 
We did the same exercise with the $V$-band  but we had to use a smaller number of intervals due to the smaller number of observations in this band. The intervals are documented in Table\,5. The long cycle usually has a smaller amplitude than in the $I$-band and the orbital light curve shows subtle variability. These changes are better visualized in the combined light curve (Fig.\,5).




\begin{figure*}
\scalebox{1}[1]{\includegraphics[angle=0,width=18cm]{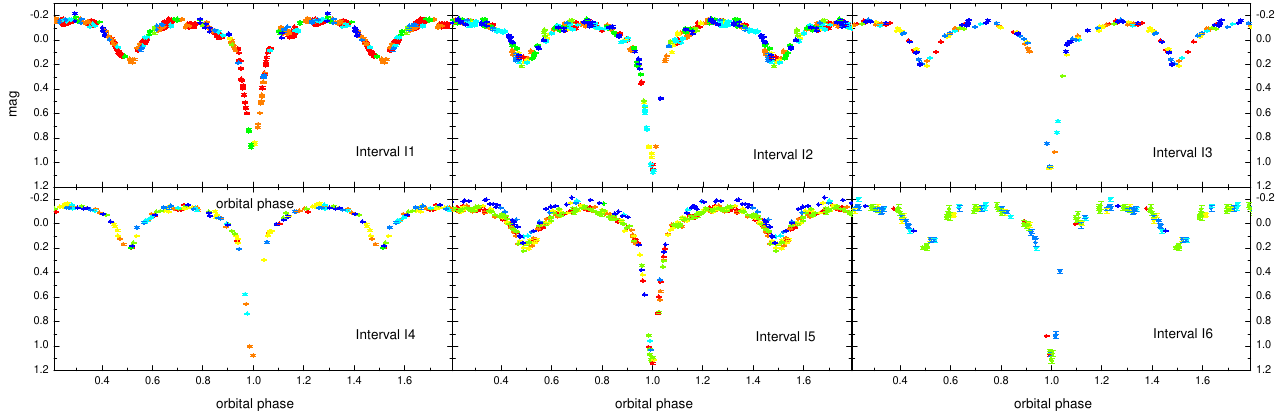}}
\scalebox{1}[1]{\includegraphics[angle=0,width=18cm]{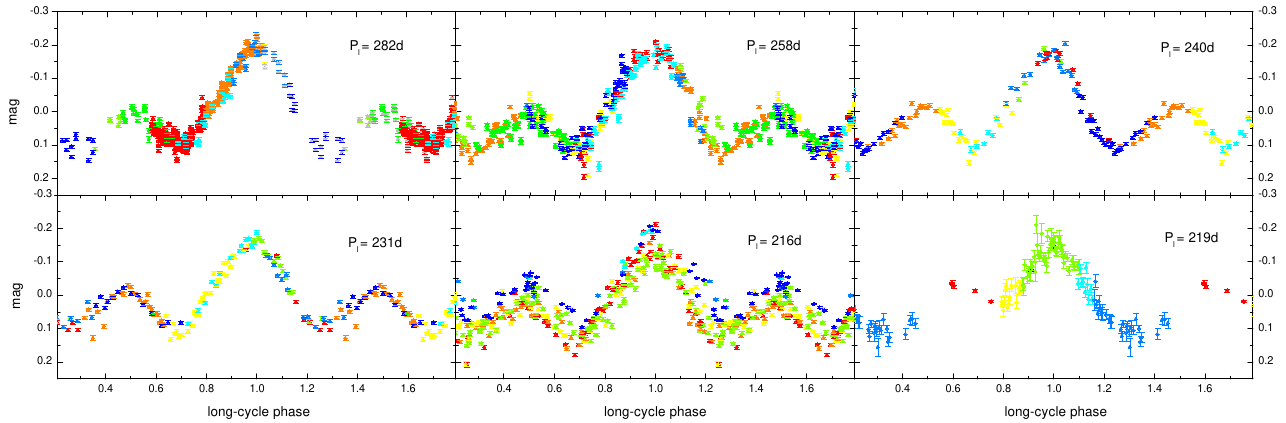}}
\caption{(up): Orbital phase curves at the intervals 1 to 6 defined in Table\,4. (Down):
Long cycle phase curves with different periods. Intervals 1 to 6 defined in Table\,4 are illustrated from the top-left to the down-right panels. In both panels 
the magnitude is differential $I$-band and the color is used to  indicate time strings of nearby data-points. }
  \label{mosaicos}
\end{figure*}

\begin{figure}
\scalebox{1}[1]{\includegraphics[angle=0,width=8cm]{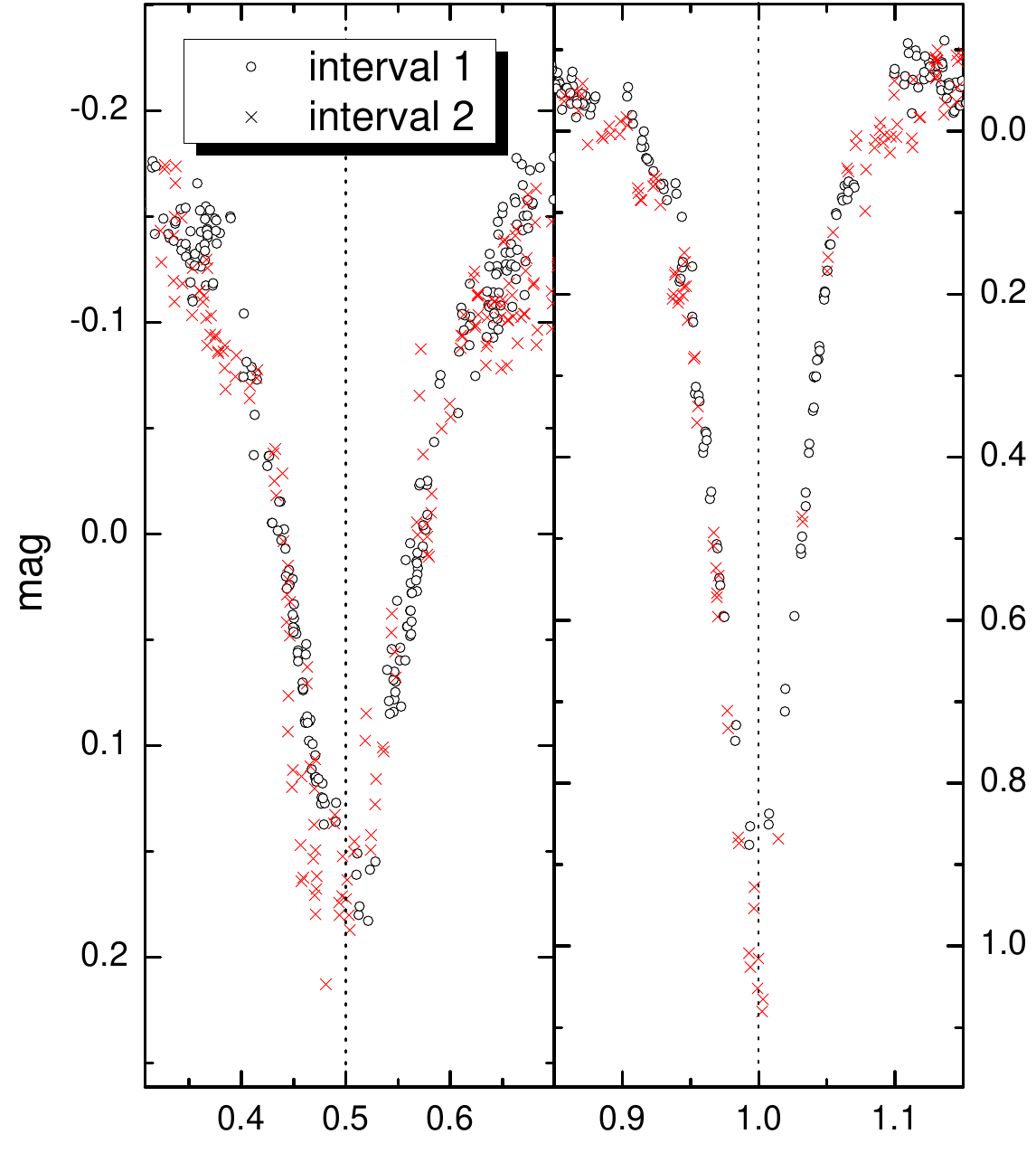}}
\scalebox{1}[1]{\includegraphics[angle=0,width=8cm]{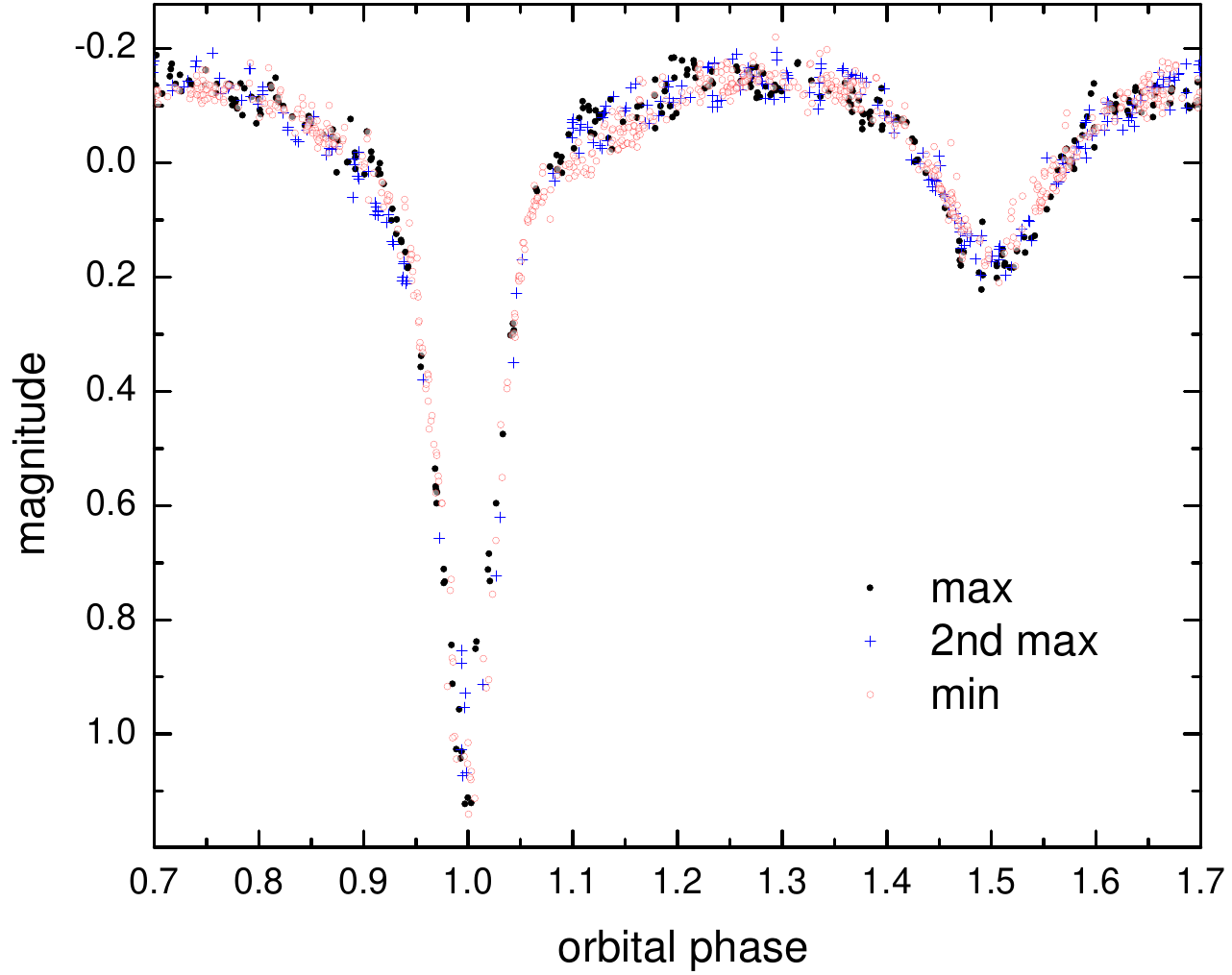}}
\caption{Up: A zoom into the secondary (left) and primary (right) eclipses for data intervals I1 and I2, as defined in Table\,4. Vertical axis shows the differential $I
$-band magnitude. Down: Eclipses during main and secondary long cycle maxima (0.9 $<$ $\Phi_l$ $\leq$ 1.1 and 0.4 $<$ $\Phi_l$ $\leq$ 0.6, respectively) and minima (
0.2 $<$ $\Phi_l$ $\leq$ 0.4 and 0.6 $<$ $\Phi_l$ $\leq$ 0.8). Vertical axis shows the differential $I$-band magnitude.}
  \label{mosaicoeclipses}
\end{figure}

\begin{figure}
\scalebox{1}[1]{\includegraphics[angle=0,width=8cm]{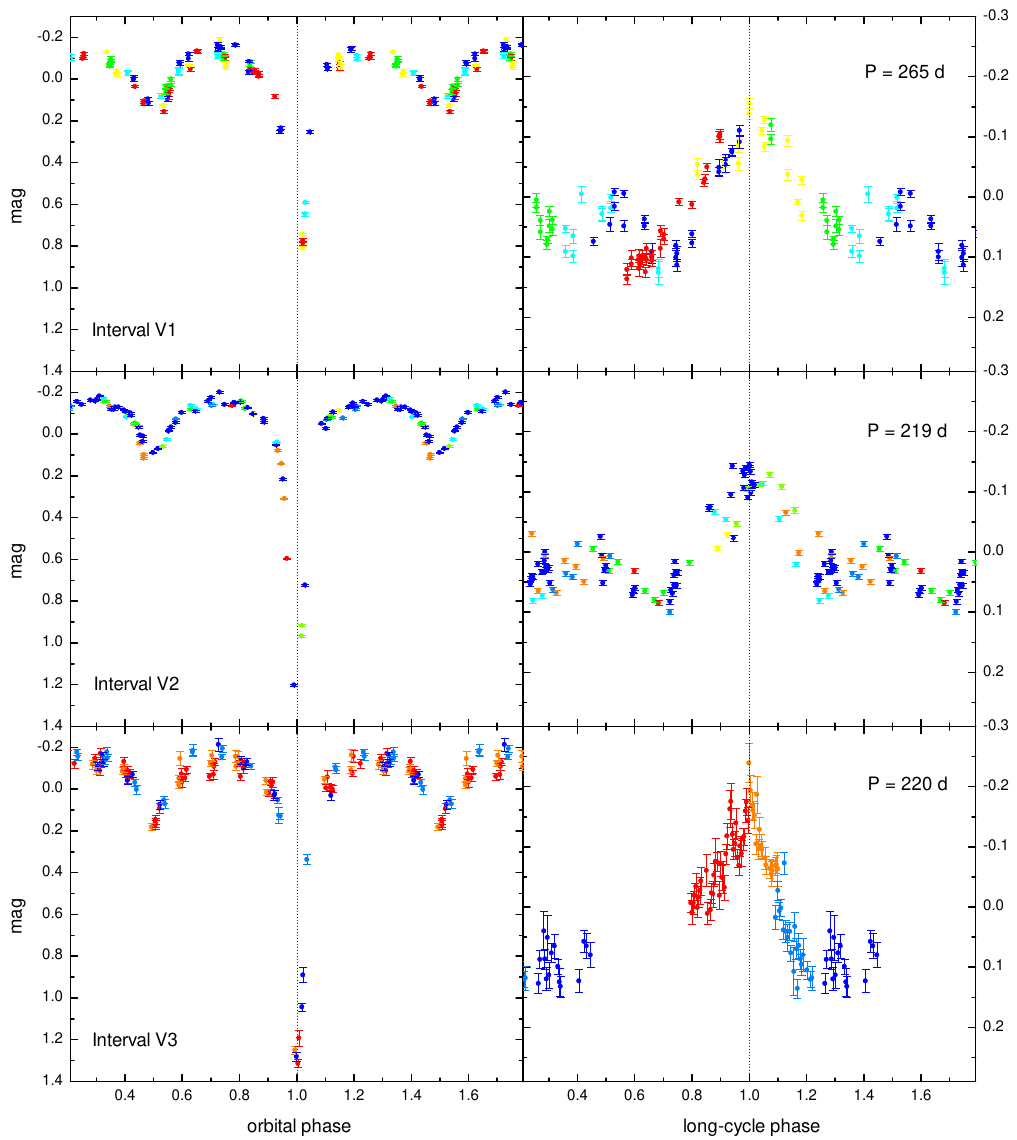}}
\scalebox{1}[1]{\includegraphics[angle=0,width=8cm]{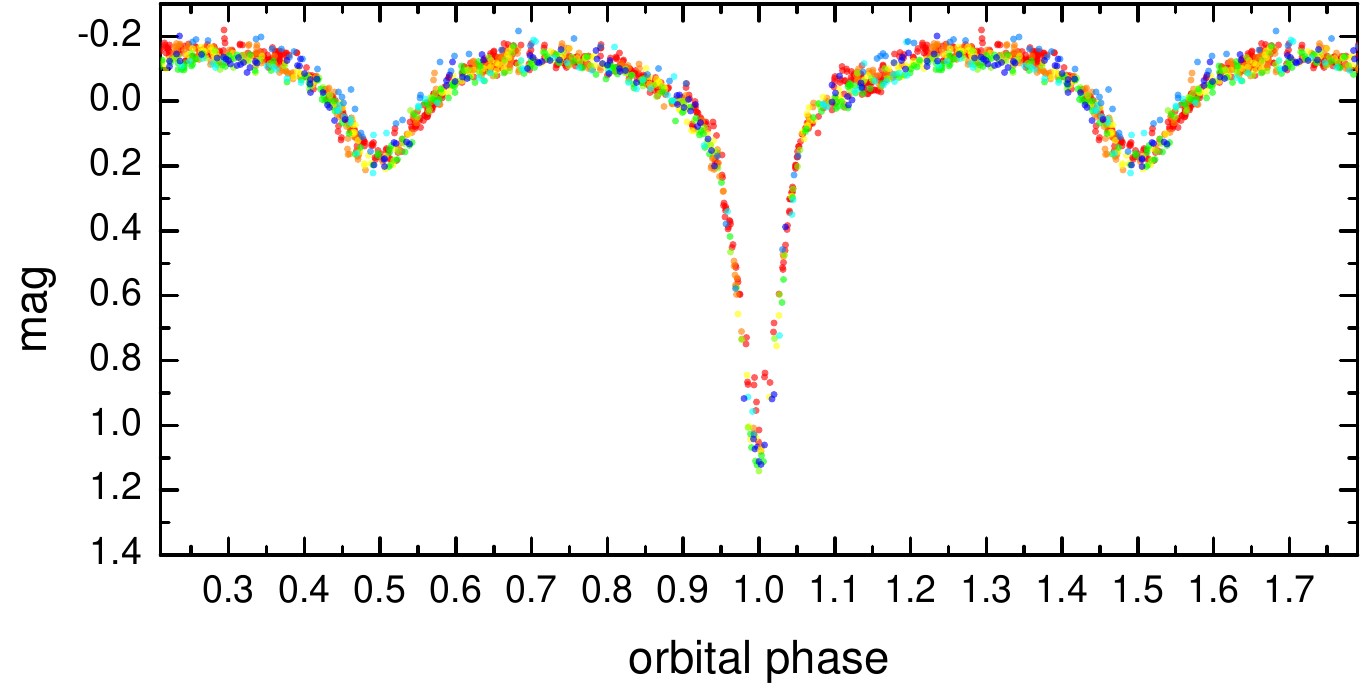}}
\caption{Up: Orbital (left) and long cycle (right)  $V$-band differential light curves at the intervals defined in Table\,5. 
The color is used to  indicate time strings of nearby data-points. Down: Combined data $V$-band light curve. Colors indicate time strings of subsequent points.}
 \label{mosaicoLCV}
\end{figure}

\begin{figure*}
\includegraphics[width=\columnwidth]{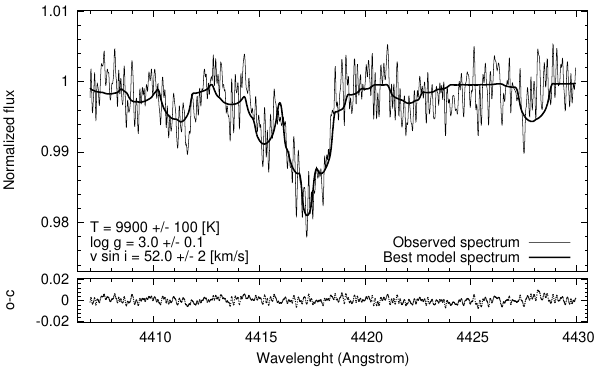}
\includegraphics[width=\columnwidth]{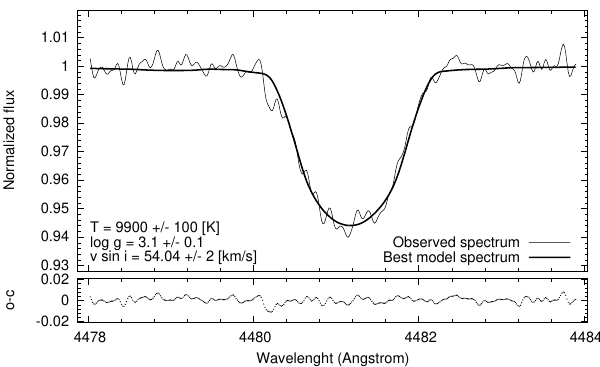}
    \caption{Disentangled donor spectrum and the best fit model at two different spectral regions.}
    \label{fig:multi}
\end{figure*}

\begin{table}
\centering
 \caption{Observed and calculated times of long cycle maxima. $N$ is the cycle number.
 HJD are referred to 2\,450\,000. $O$ stands for $HJD_{O}$ and $C$ for  $HJD_{C}$.
 Errors are given when available. We have used the linear ephemerides with zero point 2617.89 and period 240 days.
 An estimate of the local period $P_l$ is given subtracting the observed timing from the previous one  and dividing by the number of  elapsed cycles. 
}
 \begin{tabular}{@{}rrrrrc@{}}
 \hline
N &$HJD_{O}$ &$HJD_{C}$&$[O-C]$ (d) &$P_{l}$ (d) &band/source\\
\hline
-8	&571.68	$\pm$ 4.49	&  697.89&	-126.21& --       &  $I$ OGLE\\
-7	&859.75	$\pm$ 7.05	&  937.89&	-78.14&	288.06       &$I$ OGLE\\
-6	&1125.58	$\pm$ 9.48	&1177.89&	-52.31&	265.83       &$I$ OGLE\\
-4	&1645.23	$\pm$ 8.04	&1657.89&	-12.66&	259.83       &$I$ OGLE\\
-2	&2131.23	$\pm$ 10.49	&2137.89&	-6.66	    & 243.00        &$I$ OGLE\\
0	&2617.89 $\pm$ 15.47	&2617.89&	0.00	    & 243.33        &$I$ OGLE\\
2	&3077.87	$\pm$ 7.04	&3097.89&	-20.02&	229.99        &$I$ OGLE\\
3	&3311.28	$\pm$ 4.98	&3337.89&	-26.61&	233.41       &$I$ OGLE\\
5	&3747.12	$\pm$ 7.00	&3817.89&	-70.77&	217.92       &$V$ CTIO\\
5	&3749.41	$\pm$ 6.93	&3817.89&	-68.48&	219.06       &$I$ OGLE\\
6	&3973.60	$\pm$ 2.52	&4057.89&	-84.29&	224.19       &$I$ OGLE\\
8	&4403.01	$\pm$ 10.98	&4537.89&	-134.88&214.71	        &$I$ OGLE\\
8	&4408.67	$\pm$ 7.00	&4537.89&	-129.22&217.54	        &$V$ CTIO\\
10	&4835.37	$\pm$ 9.55	&5017.89&	-182.52&213.35	        &$I$ OGLE\\
13	&5477.80		        &5737.89&	-260.09	     &214.14         &$I$ OGLE\\
17	&6343.85		        &6697.89&	-354.04	     &216.51         &$y$ Turitea \\
17	&6347.25		        &6697.89&	-350.64	     &217.36         &$I$ OGLE\\
17	&6353.84	$\pm$ 8.00	&6697.89&	-344.05&219.01	        &$I$ CTIO\\
20	&7011.15	$\pm$ 4.00	&7417.89&	-406.74&219.10	       &$I$ OGLE\\
20	&7011.20	$\pm$ 2.00	&7417.89&	-406.69&219.12	       &$I$ CTIO\\
20	&7011.23	$\pm$ 10.00	&7417.89&	-406.66&219.13	       &$V$ ASAS-SN\\
22	&7448.75		        &7897.89&	-449.14	    & 218.76         &$I$ OGLE\\
22	&7453.28	$\pm$ 10.00	&7897.89&	-444.61&221.03	       &$V$ ASAS-SN\\
 \hline
\end{tabular}
\end{table}

\begin{table}
\centering
 \caption{Data intervals used for the long-period analysis. HJD are referred to 2\,450\,000.
 $n$ is the number of $I$-band data points. Times for long cycle maxima are given.
}
 \begin{tabular}{@{}rrrrr@{}}
 \hline
label &$n$ &$HJD$ range &$P_l$ (d) & $T_0 (max)$\\
\hline 
I1&538&455 $:$ 1000 &282&855.66 \\
I2&377&1000 $:$ 1900 &258&1128.82\\
I3&109&1900 $:$ 2800 &240&2618.78\\
I4&125&2800 $:$ 3500 &231&3316.75\\
I5&330&3500  $:$ 6500&216&3753.70\\
I6&104& 6500 $:$ 7500&219&7009.62\\
 \hline
\end{tabular}
\end{table}

\begin{table}
\centering
 \caption{Data intervals used for disentangling the $V$-band light curve. HJD are referred to 2\,450\,000.
 $n$ is the number of data-points. Times for long cycle maxima are given.
}
 \begin{tabular}{@{}rrrrr@{}}
 \hline
label &$n$ &$HJD$ range &$P_l$ (d) & $T_0 (max)$\\
\hline 
 V1 & 95& 468 $:$ 1632 &265  &580.73 \\
 V2 &90& 2991 $:$ 4949 &219  &2858.65 \\
 V3 &95  &6965 $:$ 7328 &220 &6789.63 \\
 \hline
\end{tabular}
\end{table}

\subsection{Spectra components and orbital/system parameters}

 In order to obtain the radial velocities and orbital parameters we
used the \texttt{KOREL}  code \citep{1995A&AS..114..393H, 1997A&AS..122..581H} 
based on the method
of Fourier disentangling, yielding directly the orbital parameters
together with the decomposed spectra of the multiple stellar system
under study. In addition, we also used the code \texttt{FOTEL}  \citep{1990CoSka..20...23H}
to estimate the errors of the orbital parameters.

We notice that the system can be classified as SB2, i.e. both
stellar components are detected in the spectrum, in particular in
helium and hydrogen lines. The detected components correspond
to an early B-type (primary or gainer), and an early A-type (secondary
or donor). The method of spectra disentangling does not use any
template or another information about the laboratory wavelengths of 
the spectral lines, therefore the systemic velocity is set to zero.
For this reason we adopted an average of systemic velocities
calculated by Gaussian adjustments for different spectral lines of
each component. For the gainer we obtain $\gamma_{pri} = 275.1 \pm 2.3$ km\,s$^{-1}$ and
for the donor $\gamma_{sec} = 279.8 \pm 2.8$ km\,s$^{-1}$. The lines used in this calculation
are shown in Table\,6. 

We notice that S II/III measurements
systematically differ from other lines, suggesting a different
formation place. For this reason they are not included in the 
above calculation.
The disentangled spectra are shown in Figs.\,6 and 7.

We performed the calculation of radial velocity in seven regions of every spectrum. 
These regions were chosen because they include several narrow, unblended and well identified metallic lines.
All our spectra were prepared with a routine written
in \texttt{IRAF6} and the sampling auxiliary code \texttt{PREKOR} \citep{2004PAICz..92...15H} 
was used. 
To diminish the numerical errors of the disentangling we sampled
each spectral region in the maximum number of bins allowed by the code, viz.\,4096. This results in the
average resolution 0.726 km\,s$^{-1}$ per bin, which is higher than
the original resolution on the spectrograph detector.

Radial velocities obtained with \texttt{KOREL} for the cases of circular orbit are  given in Table \ref{tab:rve0} and their best  fit is shown in Fig.\,8. 
 The operation of the KOREL code is described in \citet{2004PAICz..92...15H}.

The radial velocity for each components is given by
\begin{equation}
 v_j(t,p)=\sum_0 K(\cos(\omega+v)+e\cos\omega),
\end{equation}
where the sum is realized on the orbits that influence the movement of the star. The true anomaly $\upsilon$ is calculated according to
\begin{equation}
 \upsilon = 2 \arctan\left(\sqrt{\frac{1+e}{1-e}}\tan\frac{E}{2}\right),
\end{equation}
where $E$ is obtained from the solution of Kepler's equation.


 The orbital parameters obtained by disentangling of the seven spectral
regions are summarized in Table\,8. The Solution I, which we accept
for our modeling of photometry, has been obtained using an independent
disentangling of each region separately and then calculating mean solutions
and standard deviations of each parameter.
Solution II is the simultaneous ("multi-region") disentangling of all
the regions together. The errors of the parameters were obtained using
the Bayesian estimate, i.e. from the moments of the Bayesian probability
distribution \citep{2016ASSL..439..113H} We have also solved the radial-velocity
curve using the \texttt{FOTEL} code with the input radial velocities obtained
from the disentangling. The resulting values of parameters were within
the error-bars of the Solution I, but their errors were for about one
order underestimated, so we skipped this solution.
Finally, the multi-region Solution III is to verify that a possible
eccentricity of the orbit can be neglected.

\begin{table}
	\centering
	\caption{RV zero points derived from different lines. }
	\label{tab:vsys}
	\begin{tabular}{ccc} 
\hline
Spectral line& $\gamma_{\rm{pri}}$ &$\gamma_{\rm{sec}}$\\
ion $\lambda$ (\AA) &(km s$^{-1}$) &(km s$^{-1}$) \\
\hline
SiII 4128.054  &-  &281.136   \\
SiII 4130.894 &-   &278.431  \\
HeI 4143.76 &279.219 & - \\
SII 4153.068 &290.429 & - \\
NII 4227.74 & 276.360 & - \\
FeII 4233.172&- &280.476  \\  
NII 4236.91  & 277.24 &- \\
NII 4241.78  & 272.029& - \\
CrII 4242.364 &- &275.769  \\
ScII 4246.822&-&275.291\\
SIII 4253.589  & 289.218&- \\
OII 4414.884  &275.352 &-  \\
OII 4416.974  &274.255 &-  \\
HeI 4437.551 &276.146 &-  \\
TiII 4443.794 &- &277.682  \\
NII 4447.04 & 273.909 &-  \\
TiII 4533.960 &-&281.188   \\
SII 4552.410 & 289.434 &- \\
SiIII 4567.840&276.484 &-   \\
SiIII 4574.757&274.978&- \\
TiII 4549.617 &-&284.985   \\
CrII 4558.650 &-&277.747   \\
TiII 4563.757 &-&281.192   \\
TiII 4571.968 &-&282.472   \\
OII 4590.971 &274.638 &-\\
OII 4596.174 &273.442 &-\\
NII 4607.153 &272.240&- \\
SiII 4621.418 &271.289&-\\
FeII 4629.336 &-&278.677\\
OII 4638.854 &276.561 &- \\
OII 4641.811 &276.864 &- \\
OII 4649.138 &276.155 &-\\
OII 4699.21 &270.561&-\\
OII 4705.355&276.678&\\
HeI 4713.143&278.230&-\\
FeII 4731.453&-& 280.792 \\
\hline
Mean (no SII/III) &275.132$\pm$2.325&279.763$\pm$2.832\\
Mean (all) &277.030$\pm$5.464& - \\
\hline 
\end{tabular}
\end{table}

\begin{table*}
	\centering
	\caption{RVs for the primary and secondary components from \texttt{KOREL} solutions in a circular orbit. The radial velocity is the average from each spectral region and we considered the standard deviation as error.  Orbital phases $\Phi_{o}$ are given for the orbital ephemerides given by Eq.\,(1) and  long cycle phases $\Phi_{l}$ for a long period of 219 days and  $T_0 (max)$ = 57009.62.}
	\label{tab:rve0}
	\begin{tabular}{ccccrr} 
\hline															
Date-ut	&	HJD	&	$\phi_o$	&	$\phi_l$	&	$RV_g~~~~~~~$			&	$RV_d~~~~~~~~$			\\
	&	-2450000.0	&		&		&	km $s^{-1}~~~~~$			&	km $s^{-1}~~~~~$			\\
\hline															
2013-10-02 & 6567.7689 & 0.547 & 0.982 & 12.082    $\pm$    0.789 & -59.521    $\pm$    3.382    \\
2013-10-04 & 6569.8218 & 0.751 & 0.992 & 43.182    $\pm$    1.020 & -209.288$\pm$    5.953    \\
2013-10-06 & 6571.8038 & 0.949 & 0.001 & 14.536    $\pm$    0.830 & -67.676    $\pm$    2.214    \\
2013-10-07 & 6572.8075 & 0.049 & 0.005 & -13.402    $\pm$    0.618 & 61.623    $\pm$    3.562    \\
2013-10-19 & 6584.7162 & 0.236 & 0.060 & -43.399    $\pm$    1.163 & 207.660    $\pm$    7.547    \\
2013-10-22 & 6587.8479 & 0.548 & 0.074 & 12.647    $\pm$    0.685 & -60.792    $\pm$    3.658    \\
2013-12-22 & 6648.5833 & 0.603 & 0.351 & 25.914    $\pm$    0.940 & -124.089$\pm$    5.008    \\
2013-12-24 & 6650.7561 & 0.819 & 0.361 & 39.975    $\pm$    0.886 & -190.796$\pm$    4.926    \\
2013-12-31 & 6657.5987 & 0.501 & 0.393 & 0.115        $\pm$    0.649 & 0.015    $\pm$    2.346    \\
2014-01-04 & 6661.6647 & 0.907 & 0.411 & 24.468    $\pm$    0.615 & -117.341$\pm$    2.902    \\
2014-01-18 & 6675.6735 & 0.303 & 0.475 & -41.331    $\pm$    0.952 & 198.184    $\pm$    6.020    \\
2014-01-19 & 6676.6249 & 0.398 & 0.480 & -26.363    $\pm$    0.566 & 126.834    $\pm$    2.904    \\
2014-02-11 & 6699.5976 & 0.688 & 0.584 & 40.177    $\pm$    0.934 & -193.185$\pm$    5.941    \\
2014-02-15 & 6703.5569 & 0.083 & 0.603 & -21.199    $\pm$    0.864 & 103.465    $\pm$    3.704    \\
2014-02-16 & 6704.5526 & 0.182 & 0.607 & -39.697    $\pm$    0.839 & 189.726    $\pm$    6.078    \\
2014-09-01 & 6901.8820 & 0.853 & 0.508 & 34.533    $\pm$    0.659 & -169.289$\pm$    4.214    \\
2014-11-03 & 6964.8370 & 0.128 & 0.796 & -31.257    $\pm$    0.812 & 149.933    $\pm$    5.166    \\
2014-11-20 & 6981.7855 & 0.818 & 0.873 & 39.944    $\pm$    0.924 & -191.460$\pm$    5.220    \\
2014-11-22 & 6983.7327 & 0.012 & 0.882 & -3.961    $\pm$    1.684 & 13.814    $\pm$    2.907    \\
2014-11-25 & 6986.7551 & 0.313 & 0.896 & -40.347    $\pm$    0.907 & 193.721    $\pm$    5.611    \\
2014-11-26 & 6987.7523 & 0.413 & 0.900 & -23.026    $\pm$    0.504 & 111.558    $\pm$    3.302    \\
2014-11-27 & 6988.7666 & 0.514 & 0.905 & 3.562        $\pm$    0.391 & -16.488    $\pm$    3.285    \\
2014-12-08 & 6999.6148 & 0.595 & 0.954 & 24.437    $\pm$    1.068 & -116.546$\pm$    5.316    \\
2014-12-09 & 7000.7687 & 0.710 & 0.960 & 42.190    $\pm$    1.088 & -202.936$\pm$    5.340    \\
2014-12-14 & 7005.6695 & 0.199 & 0.982 & -41.140    $\pm$    0.994 & 198.112    $\pm$    6.064    \\
2015-01-20 & 7042.5709 & 0.877 & 0.151 & 30.758    $\pm$    0.632 & -147.288$\pm$    3.797    \\
2015-02-01 & 7054.5922 & 0.075 & 0.205 & -19.228    $\pm$    0.538 & 94.275    $\pm$    3.972    \\
\hline
\end{tabular}
\end{table*}

\begin{table}[htbp]
\caption{Orbital parameters obtained in different solutions}
\begin{center}
\begin{tabular}{cccc}
\hline
Parameter          &I             & II         & III\\
\hline
$P$ [d]  &\multicolumn{3}{c}{10.0316267 (fixed)}\\
$\tau$*            &92.31$\pm$0.02 &92.305$\pm$0.004&94.79$\pm0.33$ \\
$K_1$ [km\,s$^{-1}$]&42.60$\pm$0.97 &42.44$\pm$0.33  &42.45$\pm0.32$ \\
$K_2$ [km\,s$^{-1}$]&210.5$\pm$6.4  &214.1$\pm$1.8   &213.5$\pm1.7$  \\
$q$              &0.203$\pm$0.008&0.198$\pm$0.002 &0.199$\pm0.002$\\
$\omega$ [deg]         &90         &90          &178.7$\pm11.6$ \\
$e$             &0             &0              &0.021$\pm0.006$\\
\hline
\end{tabular}
\end{center}
\end{table}

\begin{figure}
\includegraphics[width=\columnwidth]{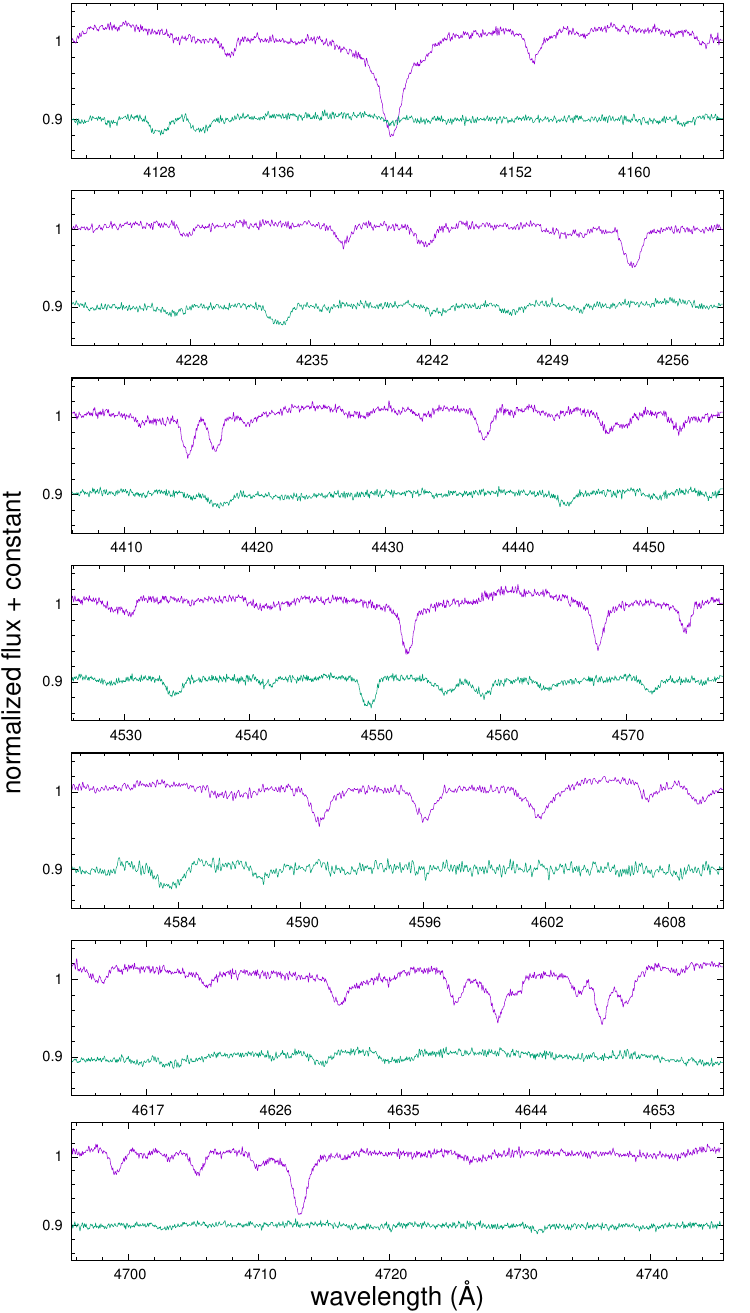}
    \caption{
  The component spectra in our seven selected regions disentangled using
 \texttt{KOREL}. For each panel, the upper spectrum corresponds to the primary
 component and the lower spectrum to the secondary component. The
 wavelength scale is shifted for the mean $\gamma$-velocity to correspond to the
 laboratory wavelengths of the lines.}
    \label{fig:multi}
\end{figure}

\begin{figure}
\includegraphics[width=\columnwidth]{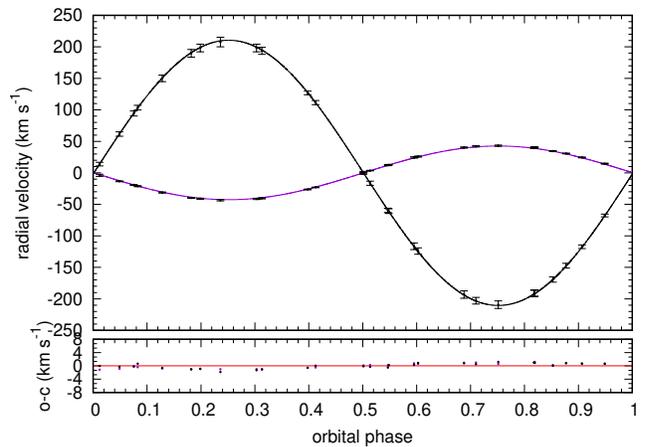}
    \caption{Fit of theoretical RVs to the average velocities from Table  \ref{tab:rve0}.}
    \label{fig:multi}
\end{figure}


Once disentangled the donor spectrum, 
we compared it with a grid of solar-metallicity synthetic models constructed with \texttt{SPECTRUM}\footnote{http://www.appstate.edu/~grayro/spectrum/spectrum.html} and 
search for the synthetic spectrum minimizing residuals.
We find the best fit with a stellar spectrum  of $T_{eff, 2}$ = 9825 $\pm$ 75 K, $v_{rot, 2}\,sin\,i$ = 53 $\pm$ 3 km s$^{-1}$ and log\,g = 3.2 $\pm$ 0.2. Comparisons of the donor disentangled spectrum with the best fit model are shown in Fig.\,6. Similarly, from the region 4120$-$4199 \AA\,  we obtained a model with 
T$_{eff, 1}$ = 22000 K and $v_{rot, 1}\,sin\,i$ = 70.6 km s$^{-1}$ for the gainer.


\section{Models for the system}

\subsection{Model for an optically thick disk around the gainer}

Part of the phenomenology of DPVs has been associated with the presence of an optically thick disk around the gainer, probably feed by a Roche-lobe filling donor star \citep[e.g.][]{2018MNRAS.477L..11G}. 
Consistently, we model the orbital light curve of OGLE-LMC-DPV-065 
considering the stellar fluxes of the two stars, the contribution of an accretion disk around the primary
and eventually the light contribution of hot/bright spots located in the outer disk rim. The basic elements of the binary system model with a plane-parallel disk and the corresponding light curve synthesis procedure are described by \citet{1992Ap&SS.196..267D, 1992Ap&SS.197.17D, 1996Ap&SS.240..317D}. The code has been successfully applied to several
close binaries including the well-studied binary system $\beta$ Lyrae
\citep[e.g.][]{2010MNRAS.409..329D, D5, 2013MNRAS.428.1594G, 2012MNRAS.421..862M, 2013MNRAS.432..799M}. 

We assume that the disk is optically and geometrically thick and that its outer edge is approximated by a cylindrical surface. The vertical thickness of the disk can change linearly with radial distance, allowing different disk's conical shapes: plane- parallel, concave or convex. The geometrical parameters of the disk are its radius ($\cal R_{\rm d}$), its vertical thickness at the outer edge (d$_e$) and the vertical thickness at the inner boundary (d$_c$). The cylindrical edge of the disk is characterized by its temperature, T$_d$, and the conical surface of the disk by a radial temperature profile inspired in the temperature distribution proposed by \citet{1991AcA....41..213Z}:\\

\begin{equation}
 \rm T(r) = T_{d} + (T_{h}+T_{d})\left[1- \left( \frac{r - \cal R_{\rm h}}{\cal R_{\rm d} - \cal R_{\rm h}} \right) \right]^{a_{T}}
\end{equation} 

We further assume that the disk is in physical contact and thermal equilibrium with the gainer, so its inner radius and corresponding temperature are equal to the radius and temperature of the star ($\cal R_{\rm h}$, T$_h$). The temperature of the disk at the edge (T$_d$) and the temperature exponent (a$_T$), as well as the radii of the gainer ($\cal R_{\rm h}$) and of the disk ($\cal R_{\rm d}$) are free parameters, determined by solving the inverse problem (see Section 4.2).

Motivated by previous research on DPVs (Mennickent et al. 2016), our model includes active regions on the edge of the disk. These active regions are usually revealed in Doppler tomography maps of disks in Algol  binaries \citep{2004AN....325..229R}. 
These regions have higher local temperatures than the disk, and produce a non-uniform distribution of radiation. We consider two active regions: a hot spot (h$_s$) and a bright spot (b$_s$), characterized by their temperatures T$_{hs, bs}$, angular dimensions (radius) $\theta_{hs,bs}$ and longitudes $\lambda_{hs,bs}$ . The longitude $\lambda$ is measured from the line joining the gainer and donor centers in the direction opposite to the orbital motion in the orbital plane. These parameters are also determined by solving the inverse problem. We also consider a possible departure of symmetry of light emerging from the hot spot due, for instance, to the impact of the gas stream in the disk.
This deviation is described by the angle $\theta_{rad}$ between the line perpendicular to the local disk edge surface and the direction of the hot spot maximum radiation in the orbital plane.
The second spot in the model (here named bright spot), simulates the spiral structure of an accretion disk, observed in hydrodynamical calculations \citep{1994A&A...288..807H}. The tidal forces exerted by the donor star produce a spiral shock, observed as one or two extended spiral arms in the outer disk regions. This bright spot can also be interpreted as a region where the disk significantly deviates from the circular shape.

Two potential limitations of the code need to be briefly mentioned:  the lack of treatment of the donor irradiation by the hot spot, and the lack of inclusion of a possible not eclipsed {\it  additional} third light, considering that the long-cycle light was already removed with the process of  disentangling. However, the very good fit obtained (based on $\chi^2$ minimization) suggests that these additional light sources, if present, are much fainter than the stars and the disk/spots.  In addition, while the donor irradiation by the hot spot is not included, the much larger effect of the donor irradiation by the gainer is implemented in our code.

\subsection{The fit to the orbital light curve}

The fit to the orbital light curve was performed using the inverse-problem solving method based on the simplex algorithm, and the model for the binary system described in the previous section. The inverse-problem method is the process of finding the set of parameters that will optimally fit the synthetic light curve to the observed data. We used the Nelder-Mead simplex algorithm \citep{1992nrfa.book.....P} with the optimization described by Dennis \& Torczon (1991).  For details see \citet{1992Ap&SS.197.17D}. 

Based on results of the previous sections we fixed the spectroscopic mass ratio to $q$ = 0.203 and the donor temperature to T$_2$ = 9825 K. In addition, we set the gravity darkening coefficient and the albedo of the gainer and the donor to $\beta_{h,c}$  = 0.25 and A$_{h,c}$ = 1.0, following von Zeipel's law for radiative shells and complete re-radiation \citep{1924MNRAS..84..702V}. The limb darkening for the components was calculated as described by \citet{2010MNRAS.409..329D}.




We assume that the donor is rotating synchronously, i.e. the non-synchronous rotation coefficient, defined as the ratio between the actual and the Keplerian angular velocity is f$_c$ = 1.0. This is justified since it is assumed that the donor has filled its Roche lobe (i.e. the filling factor of the donor was set to F$_c$ = 1.0), then the synchronization is expected as consequence of the tidal forces.

The case for the gainer is different, since the accreted material from the disk is expected to transfer angular momentum increasing the rotational speed of the gainer up to the critical velocity as soon as even a small fraction of the mass has been transferred \citep{1981A&A...102...17P, 2007AIPC..948..321D, 2013A&A...557A..40D}.
For this reason we assumed critical rotation for the gainer, and estimated a non-synchronous rotation factor f$_h$ = 8.9 in the critical rotation regime. 



The best fit along with the $O-C$ residuals, individual donor, disk and gainer flux contributions and the view of the optimal model at orbital phases 0.25, 0.50 and 0.75, are shown in Fig.\,9. We note that the residuals show no dependence on orbital or long-cycle phases, except a larger random scatter around main eclipse.  Parameters are given in Table\,9 and the sensitivity of $\chi ^2$ with some parameters is illustrated in Fig.\,10.
We find that at quadrature and $I$-band, the gainer contributes 27\% more flux than the donor and the disk only 48\% of the donor to the  total flux.

We find that the system contains a 13.8  \msun\ star and a 2.81 \msun\ star with absolute magnitude $M_{bol}$ = -6.4 and -3.0 respectively, separated by 49.9 \rsun. The stellar temperatures are T$_h$ = 25460 K and T$_c$  (fixed) = 9825 K.
The best-fitting model contains an optically and geometrically thick disk around the hotter, more massive gainer star. With a radius of  $\cal R_{\rm d}$
  $\approx$  25 \rsun, the disk is 2.8 times larger than the central star ($\cal R_{\rm h}$ $\approx$ 8.8 \rsun). The disk has a convex shape, with central thickness d$_c$ $\approx$ 1.6 \rsun\ and edge thickness d$_e$ $\approx$ 3.5 \rsun. The temperature of the disk decreases from T$_d$ = 9380 K at its edge, to T$_h$ = 25460 K at the inner radius.

We notice that the hot spot has 20\fdg5 angular radius and covers 12\% of the disk outer rim, and it is situated at longitude $\lambda_{hs}$ = 312\fdg4, roughly between the components of the system, at the place where the gas stream falls on to the disk \citep{1975ApJ...198..383L}. The temperature of the hot spot is approximately 18 per cent higher than the disk edge temperature, i.e. T$_{hs}$ =  11068 K. Although including the hot spot region into the model improves the fit, it cannot explain the light curve asymmetry completely. By introducing one additional bright spot, larger than the hot spot and located on the disk edge at $\lambda_{bs}$ = 114\fdg9, the fit becomes much better. This bright spot is 26\% hotter than the disk at its edge, i.e. T$_{bs}$ =  11819 K and has an angular radius of 27\fdg9,  covering 16\%  of the disk outer rim.


\section{Discussion}


Only a few DPVs have been studied spectroscopically in detail and therefore few of them posses relatively well-determined orbital and stellar parameters;
9 Galactic DPVs and the LMC DPV OGLE05155332-6925581 are documented by \citet{2016MNRAS.455.1728M} and recently  stellar and orbital parameters were provided for V\,495\,Cen by \citet{2018MNRAS.476.3039R}.
Our study of OGLE-LMC-DPV-065 presented in this paper is the second spectroscopic study of an LMC DPV.

In  Fig.\,11 we compare OGLE-LMC-DPV-065 data with those of other DPVs and classical Algols, these later taken as reference. It is clear that DPVs are 
hotter and more massive than ordinary Algols, a fact already noticed in previous studies. In addition, 
it is clear that OGLE-LMC-DPV-065 is a comparatively massive and hot DPV, in many aspects similar to $\beta$ Lyrae.    
In Table\,10 we provide a comparison between these systems based on the results of  \citet{2013MNRAS.432..799M}, 
although see also the recent research on the $\beta$ Lyrae disk by \citet{2018A&A...618A.112M} for a complementary approach 
 confirming the existence of a hot spot and obtaining roughly the same disk size but from an interferometric study.

The similarity is especially significant in inclination angle, stellar masses, surface gravities and time scale of the long-cycle length. Both systems are found in a mass transfer stage,  harbor a comparatively hot accretion disk  and massive B $+$ A type stars for the DPV standard (Fig.\,9). The radial extension of the disk is also similar along with the location of the hot and bright spots. As a jet has been detected in $\beta$ Lyrae  \citep{1996A&A...312..879H, 2007A&A...463..233A, 2011BSRSL..80..689L}, it is then possible that the same structure exists in OGLE-LMC-DPV-065 and could be related to the long-cycle through a magnetic dynamo as suggested by \citet{2017A&A...602A.109S}. On the other hand, an important difference is the remarkable long-cycle change observed in OGLE-LMC-DPV-065 which is not observed in $\beta$ Lyrae. The large amplitude of the long-cycle in OGLE-LMC-DPV-065 is also remarkable. In comparison, the long-cycle in $\beta$ Lyrae is of low amplitude and relatively constant in period. 
Orbital period changes can be explained in terms of conservative mass transfer in a binary system.
Hence it is possible that both systems are in different stages of the mass transfer episode. 
A much larger mass transfer in $\beta$ Lyrae might explain why this binary shows a variable orbital period, whereas OGLE-LMC-DPV-065, eventually with a smaller mass transfer rate, does not. In addition, 
$\beta$ Lyrae has a larger and brighter secondary star, which might also play a role in the observed differences between both systems.  These issues will be investigated in a forthcoming paper.


If a magnetic dynamo is the cause for the long-cycle, then these two systems with similar parameters but different long cycle light curve morphology, constitute constrains to be satisfied by any competent detailed physical model of the long variability. Our next study will explore this point, establishing the evolutionary stage of OGLE-LMC-DPV-065 and analyzing the spectroscopic changes during the long cycle.
We will also present numerical calculations  aimed to test the hypothesis of variable mass transfer driven by a magnetic dynamo 
as proposed by \citet{2017A&A...602A.109S}.


 \section{Conclusions}

We have analyzed the variability of the eclipsing Algol OGLE-LMC-DPV-065 considering new and published photometric data spanning 124 years. The orbital and long-cycle light curves
have been disentangled and characterized. We also presented the first spectroscopic study of this binary system obtaining the mass ratio and temperature of the cooler stellar component. These quantities served as fixed input parameters in our model of the light curve, that was done following an inverse-problem methodology. The best solution shows a reasonable fit to the light curve providing additional parameters for the binary, the stellar components and the circumprimary accretion disk.
The main results of our research can be summarized as follows:\\

\begin{itemize}
\item We find a refined orbital period of 10\fd0316267 $\pm$ 0\fd0000056 without any evidence of variability.
\item Small but significant changes in timings of the secondary eclipse are detected. They might be caused by circumstellar material.
\item The long-cycle is characterized by a double hump light-curve at $I$ and $V$ bands, of amplitude about 0.3 and 0.2 mag, respectively, 
whose general shape is more or less constant, with only minor variability.
\item  We find that after a  continuous decrease of the long-period during about  13 years, from 350 to 218 days, it remained almost constant by about 10 years.
\item The study of radial velocities indicates a binary in a circular orbit with mass ratio of 0.203  $\pm$ 0.001.
\item We find that the system consists of a pair of stars of 13.8 and 2.81 \msun\ of radii 8.8 and 12.6 \rsun\ and absolute bolometric magnitudes -6.4 and -3.0, respectively. 
\item We find stellar temperatures of 25460 K and 9825 K for the gainer and donor, respectively. 
\item We find an orbital semi-major axis of 49.9 \rsun.
\item We find evidence of an accretion disk with a radius of  25 \rsun, central thickness 1.6 \rsun\ and edge thickness 3.5 \rsun.  The temperature of the disk decreases from 25460 K at the inner radius to 9380 K at its outer edge.
\item As happens in other DPVs, two hot shock regions located at roughly opposite parts of the outer disk rim can explain the light curves asymmetries.  
\item OGLE-LMC-DPV-065 resembles in some aspects to the well-studied binary $\beta$ Lyrae. However, its orbital period 
does not change, this could indicate a smaller mass transfer rate.
\end{itemize}

\begin{table}

\caption{Results of the analysis of {DPV065} light-curves
obtained by solving the inverse problem for the Roche model with
an large accretion disk partially obscuring the more-massive (hotter) gainer in critical non-synchronous
rotation regime.}
\label{DPV065}
\[
\begin{array}{@{\extracolsep{+0.0mm}}llllllll@{}}
\hline
\noalign{\smallskip}
{\rm Quantity} & & & & {\rm Quantity} \\
\noalign{\smallskip}
\hline
\noalign{\smallskip}
n 					& 718	 & & & & & & \\
{\rm \Sigma(O-C)^2}			& 0.4842 & & & & & & \\
{\rm \sigma_{rms}}			& 0.0260 & & & & & & \\
i {\rm [^{\circ}]}          & 86.7 \pm 0.4  & &  & \cal M_{\rm_h} {[\cal M_{\odot}]} & &	& 13.8 \pm 0.3 \\
{\rm F_d}				& 0.948 \pm 0.03 & &  & \cal M_{\rm_c} {[\cal M_{\odot}]} & &	& 2.81 \pm 0.2 \\
{\rm T_d} [{\rm K}]			&  9380 \pm 400 & &  & \cal R_{\rm_h} {\rm [R_{\odot}]} & &	& 8.8 \pm 0.3 \\
{\rm d_e} [a_{\rm orb}]			& 0.071 \pm 0.005 & &  & \cal R_{\rm_c} {\rm [R_{\odot}]} & & 	& 12.6 \pm 0.3 \\
{\rm d_c} [a_{\rm orb}]			& 0.032 \pm 0.008 & &  & {\rm log} \ g_{\rm_h}		& &	& 3.68 \pm 0.05 \\
{\rm a_T}				& 4.6 \pm 0.5   & &  & {\rm log} \ g_{\rm_c}	& &		& 2.69 \pm 0.06 \\
{\rm F_h}				& 1.000  & &  & M^{\rm h}_{\rm bol}	& &		& -6.39 \pm 0.2 \\
{\rm T_h} [{\rm K}]			& 25460 \pm 1500 & &  & M^{\rm c}_{\rm bol}		& &	& -3.01 \pm 0.1 \\
{\rm T_c} [{\rm K}]			& 9825   & &  & a_{\rm orb}  {\rm [R_{\odot}]}  & &     & 49.91 \pm 0.2\\
{\rm A_{hs}=T_{hs}/T_d}			&  1.18 \pm 0.07 & &  & \cal{R}_{\rm d} {\rm [R_{\odot}]} & &	& 25.0 \pm 0.4 \\
{\rm \theta_{hs}}{\rm [^{\circ}]}	&  20.5 \pm 0.6 & &  & \rm{d_e}  {\rm [R_{\odot}]}	& &	& 3.5 \pm 0.2  \\
{\rm \lambda_{hs}}{\rm [^{\circ}]} 	& 312.4 \pm 8.0 & &  & \rm{d_c}  {\rm [R_{\odot}]}	& &	& 1.6 \pm 0.1  \\
{\rm \theta_{rad}}{\rm [^{\circ}]} 	& -21.5 \pm 2.2 & &  & & & & \\
{\rm A_{bs1}=T_{bs}/T_d}		&  1.26 \pm 0.07 & &  & & & & \\
{\rm \theta_{bs}} {\rm [^{\circ}]}	&  27.9 \pm 4.0 & &  & & & & \\
{\rm \lambda_{bs}}{\rm [^{\circ}]}	& 114.9 \pm 10.0 & &  & & & & \\
{\rm f_h}				& 8.9 \pm 0.3   & &  & & & & \\
{\rm \Omega_h}				& 7.067 \pm 0.02 & &  & & & & \\
{\rm \Omega_c}				& 2.240 \pm 0.02 & &  & & & & \\
\noalign{\smallskip} \hline \noalign{\smallskip}

\noalign{\smallskip}
\hline
\end{array}
\]
\bigskip

FIXED PARAMETERS: $q={\cal M}_{\rm c}/{\cal M}_{\rm h}=0.203$ - mass ratio of
the components, ${\rm T_c=9825 K}$  - temperature of the less-massive (cooler)
donor,
${\rm F_c}=1.0$ - filling factor for the critical Roche lobe of the donor,
$f{\rm _h}=8.9 ; f{\rm _c}=1.00$ - non-synchronous rotation coefficients of
the gainer and donor respectively, ${\rm \beta_h=0.25 ; \beta_c=0.25}$ -
gravity-darkening coefficients of the gainer and donor, ${\rm A_h=1.0 ;
A_c=1.0 ; A_d=1.0}$  - albedo coefficients of the gainer, donor and disk.

\smallskip \noindent Quantities: $n$ - number of observations,
${\rm \Sigma (O-C)^2}$ - final sum of squares of residuals between observed
(LCO) and synthetic (LCC) light-curves, ${\rm \sigma_{rms}}$ - root-mean-square
of the residuals, $i$ - orbit inclination (in arc degrees),
${\rm F_d=R_d/R_{yc}}$ - disk dimension factor (ratio of the disk radius to
the critical Roche lobe radius along y-axis), ${\rm T_d}$ - disk-edge
temperature, $\rm{d_e}$, $\rm{d_c}$,  - disk thicknesses (at the edge and at
the center of the disk, respectively) in the units of the distance between the
components, $a_{\rm T}$ - disk temperature distribution coefficient,
${\rm F_h}=R_h/R_{zc}$ - filling factor for the critical Roche lobe of the
hotter, more-massive gainer (ratio of the stellar polar radius to the
critical Roche lobe radius along z-axis  for a star in critical rotation regime),
${\rm T_h}$ - temperature of the
more-massive (hotter) gainerr, ${\rm A_{hs,bs}=T_{hs,bs}/T_d}$ - hot and bright
spots' temperature coefficients, ${\rm \theta_{hs,bs}}$ and ${\rm \lambda_{hs,bs}}$ -
spots' angular dimensions and longitudes (in arc degrees), ${\rm \theta_{rad}}$
- angle between the line perpendicular to the local disk edge surface and the
direction of the hot-spot maximum radiation, $f{\rm _h}$- non-synchronous rotation
coefficients of the gainer in critical rotation regime, ${\rm \Omega_{h,c}}$ -
dimensionless surface potentials of the hotter gainer and cooler donor,
$\cal M_{\rm_{h,c}} {[\cal M_{\odot}]}$, $\cal R_{\rm_{h,c}} {\rm [R_{\odot}]}$
- stellar masses and mean radii of stars in solar units,
${\rm log} \ g_{\rm_{h,c}}$ - logarithm (base 10) of the system components
effective gravity, $M^{\rm {h,c}}_{\rm bol}$ - absolute stellar bolometric
magnitudes, $a_{\rm orb}$ ${\rm [R_{\odot}]}$, $\cal{R}_{\rm d} {\rm [R_{\odot}]}$,
$\rm{d_e} {\rm [R_{\odot}]}$, $\rm{d_c} {\rm [R_{\odot}]}$ - orbital semi-major axis,
disk radius and disk thicknesses at its edge and center, respectively, given in the
solar radius units.

\end{table}

\begin{figure}
\includegraphics[width=5cm]{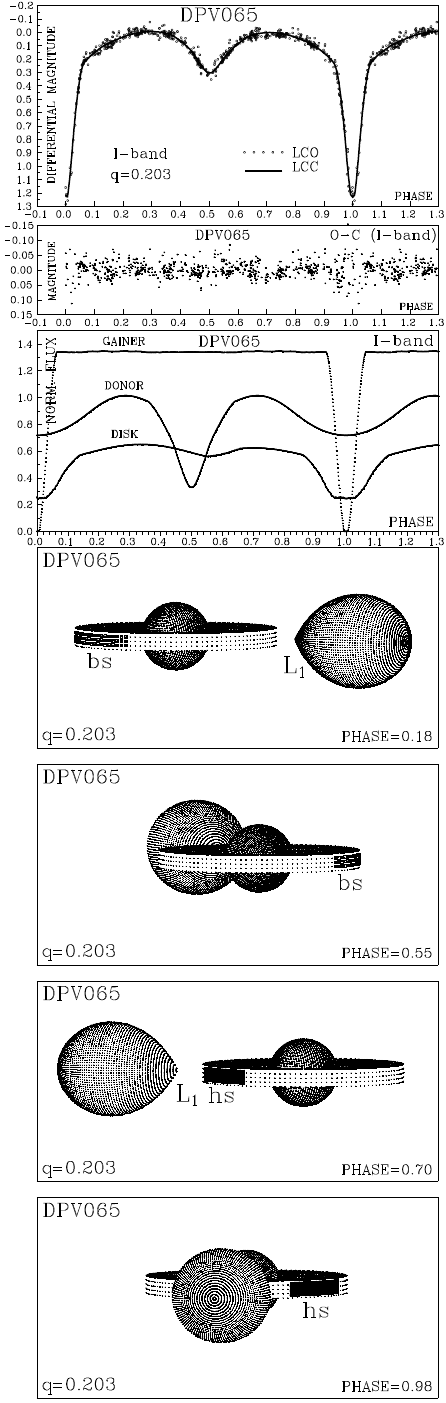}
\caption{Observed (LCO), synthetic (LCC) light-curves and the final
O-C residuals between the observed and synthetic light curves of
OGLE-LMC-DPV-065; fluxes of donor, gainer and of the accretion disk,
normalized to the donor flux at phase 0.25; the views
of the model at orbital phases 0.18, 0.55, 0.70 and 0.98,
obtained with parameters estimated by the light curve analysis.  This is for a gainer in critical non-synchronous
rotation regime.}
\label{fDPV065}
\end{figure}

\begin{figure}
\includegraphics[width=\columnwidth]{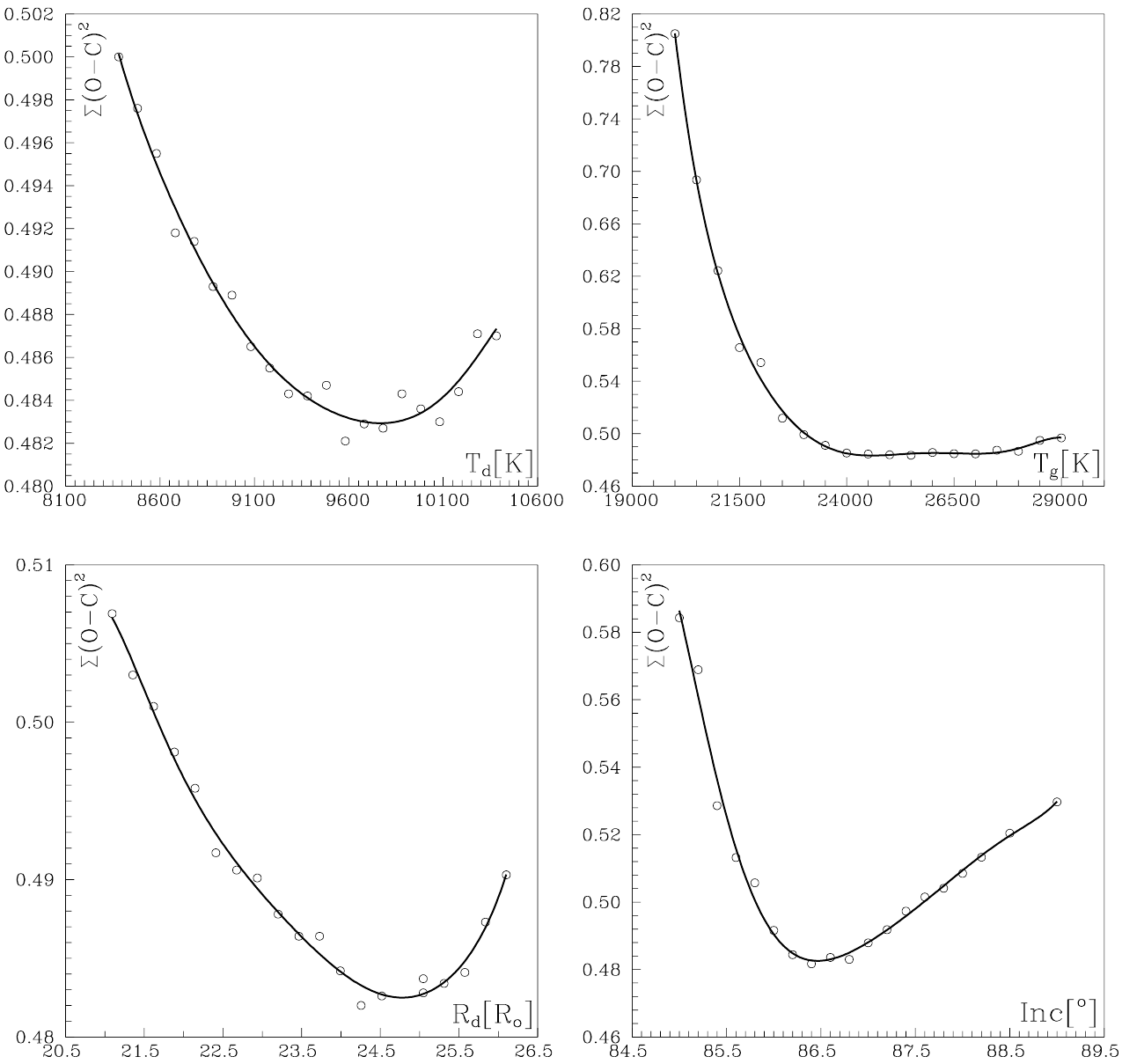}
    \caption{ Plots showing the dependence of \bf $\chi ^2$.}
    \label{fSfun}
\end{figure}

\begin{figure}
\includegraphics[width=\columnwidth]{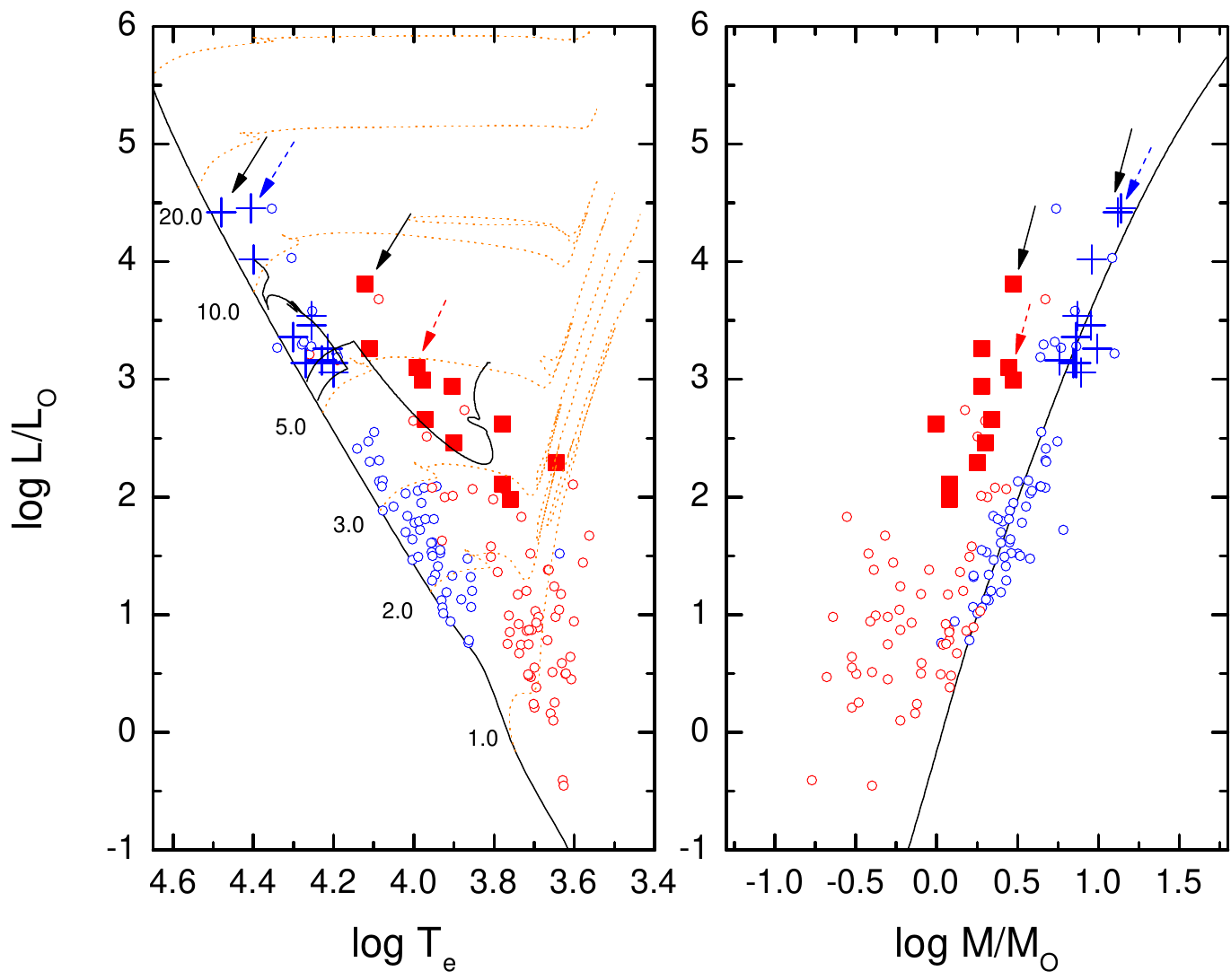}
    \caption{Comparison of physical data for semi-detached Algols from \citet[][primaries open blue circles and secondaries open red circles]{2010MNRAS.406.1071D} and DPVs
\citep[][(primaries blue crosses and secondaries red squares)]{2016MNRAS.455.1728M}. The zero-age main sequence for Z = 0.02 is plotted with a solid black line and evolutionary tracks for single stars with initial masses (in solar masses) labelled at the track footprints are also shown \citep{1998MNRAS.298..525P}. The positions of the gainer and donor of OGLE-LMC-DPV-065 are indicated by  color arrows, whereas $\beta$ Lyrae is labeled with black arrows. The best evolutionary tracks for the primary and secondary of the DPV HD\,170582 are also plotted by two solid lines in the left-hand panel \citep{2016MNRAS.455.1728M}.}
    \label{plotLMCDPV}
\end{figure}

\section*{Acknowledgments}

 We thanks the referee, Denis Mourard, who helped to improve the first version of this manuscript. This paper uses photometric data acquired under CNTAC proposal CN2014B-13.
This research has made use of the SIMBAD database, operated at CDS, Strasbourg, France.
R.E.M. gratefully acknowledges support by VRID-Enlace 218.016.004-1.0, FONDECYT 1190621, 
and the Chilean Centro de Excelencia en Astrof{\'{i}}sica
y Tecnolog{\'{i}}as Afines (CATA) BASAL grant AFB-170002. 
The OGLE project has received funding from the Polish National Science
Centre grant MAESTRO no. 2014/14/A/ST9/00121. G. D. gratefully acknowledges the financial support of
the Ministry of Education and Science of the Republic of Serbia
through the project 176004, Stellar physics. J.G.F.-T. is supported by FONDECYT N. 3180210.
N.A-D. acknowledges support from FONDECYT \#3180063.
We thanks Shelby Owens for reducing data of the Turitea Observatory.

\begin{table}

\caption{Comparison between the  OGLE-LMC-DPV-065 data obtained in this paper and those of  $\beta$ Lyrae obtained by \citet{2013MNRAS.432..799M} and references therein.}
\label{comp}
\[
\begin{array}{@{\extracolsep{+0.0mm}}lrr@{}}
\hline
\noalign{\smallskip}
{\rm Quantity} &{\rm OGLE-LMC-DPV-065} & {\rm beta\,Lyrae}  \\
\noalign{\smallskip}
\hline
\noalign{\smallskip}
\hline
{\rm P_o}  [{\rm days}]                 &10.0316267 &12.95  {\rm (variable)} \\
{\rm P_l}  [{\rm days}]                 &350-210 {\rm (variable)} &282.4 \\
{\rm P_l/P_o}  &35-21 &22 \\ 
i {\rm [^{\circ}]}                             & 86.7  &86.1 \\
{\rm T_d} [{\rm K}]	                   & 9380 & 8200\\
{\rm d_e} [a_{\rm orb}]		  &0.071&0.192  \\
{\rm d_c} [a_{\rm orb}]		  &0.032&0.01  \\
{\rm a_T}				          & 4.6       &3.8 \\
{\rm T_h} [{\rm K}]			  & 25460&30000 \\
{\rm T_c} [{\rm K}]			  & 9825&13300 \\
{\rm A_{hs}=T_{hs}/T_d}		 &1.18&1.21 \\
{\rm \lambda_{hs}}{\rm [^{\circ}]} &312.4&324.6 \\
{\rm A_{bs}=T_{bs}/T_d}		 &1.26&1.12  \\
{\rm \lambda_{bs}}{\rm [^{\circ}]} &114.9&107.3\\
\cal M_{\rm_h} {[\cal M_{\odot}]} & 13.8 & 13.16\\
\cal M_{\rm_c} {[\cal M_{\odot}]} & 2.81 &2.97 \\
\cal R_{\rm_h} {\rm [R_{\odot}]} & 8.8   &6.0\\
\cal R_{\rm_c} {\rm [R_{\odot}]} & 12.6& 15.2 \\
 {\rm log} \ g_{\rm_h}			& 3.68& 4.0 \\
 {\rm log} \ g_{\rm_c}			& 2.69  &2.5 \\
 M^{\rm h}_{\rm bol}			& -6.39 &-6.3\\
M^{\rm c}_{\rm bol}			& -3.01& -4.7\\
a_{\rm orb}  {\rm [R_{\odot}]}    & 49.91&58.5 \\
\cal{R}_{\rm d} {\rm [R_{\odot}]} & 25.0& 28.3 \\
\rm{d_e}  {\rm [R_{\odot}]}		& 3.5  &11.2 \\
 \rm{d_c}  {\rm [R_{\odot}]}		& 1.6  &0.6  \\
\hline
\noalign{\smallskip}
\hline
\end{array}
\]
\end{table}

\bsp 
\label{lastpage}

\newpage

\end{document}